\documentstyle[aps,multicol,epsfig,prb]{revtex}
\newcommand{\blong}{\ifpreprintsty
                   \else
                   \end{multicols}\vspace*{-3.5ex}{\tiny
                   \noindent\begin{tabular}[t]{c|}
                   \parbox{0.49\hsize}{~} \\ \hline \end{tabular}}
                   \fi}
\newcommand{\elong}{\ifpreprintsty
                   \else
                   {\tiny\hspace*{\fill}\begin{tabular}[t]{|c}\hline
                    \parbox{0.49\hsize}{~} \\
                    \end{tabular}}\vspace*{-2.5ex}\begin{multicols}{2}
                    \fi}
\def\be{\begin{equation}}
\def\ee{\end{equation}}
\def\bea{\begin{eqnarray}}
\def\eea{\end{eqnarray}}

\def\Bi2212{Bi$_{2}$Sr$_{2}$CaCu$_{2}$O$_{8+\delta}$}
\def\Hc{{\cal H}_{c}}
\def\ie{ {\it i.e.} }
\def\etal{ {\it et.al.} }
\begin{document}
\title{$C$-axis Optical Sum Rule in Josephson Coupled Vortex State}
\author{Wonkee Kim and J. P. Carbotte}
\address{Department of Physics and Astronomy, McMaster University,
Hamilton, Ontario, Canada L8S~4M1}
\maketitle
\begin{abstract}
Observed violations of the $c$-axis optical sum rule can give important
information on deviations from in-plane Fermi liquid behavior and on
the nature of interlayer coupling between adjacent copper
oxide planes. Application
of a magnetic field perpendicular to these planes is another way to
probe in-plane dynamics. We find that the optical sum rule
is considerably modified in the presence of the $c$-axis magnetic field.
Interlayer correlation of pancake vortices is
involved in the sum rule modification; however, details of 
the vortex distribution
in the plane are less important.
\end{abstract}
\begin{multicols}{2}
\tighten

\section{introduction}

The conductivity sum rule is one of the most useful methods to analyze
optical properties of high-$T_{c}$ cuprates. 
The conventional sum rule\cite{fgt} states that the missing
spectral weight $\Delta N$ under the real part of the optical 
conductivity between superconducting and normal state is equal to the 
superfluid density $\rho_{s}$.
However, violations of the conventional sum rule along the $c$-axis
have been observed\cite{basov} in some high-$T_{c}$
cuprates, and are related to the change in kinetic energy
\cite{hirsh,ioffe,chakravarty,kim1}
on entering the superconducting state.
Since $c$-axis response reflects
in-plane dynamics,\cite{kim1} 
the $c$-axis conductivity sum rule and the corresponding
superfluid density are worthwhile investigating to understand characteristics
of CuO$_{2}$ planes as well as for their own importance and interest. 

We have previously studied\cite{kim2} 
effects of an in-plane magnetic field on the $c$-axis sum rule
and on the superfluid density under the assumption that the in-plane field
freely penetrates between CuO$_{2}$ planes, 
and found that such a field
could not easily change the $c$-axis sum rule.
The reason for this is that the in-plane field is 
not directly pair breaking and it is only the induced vortex due to Josephson
tunneling that reduces the superfluid density. 
Usually the conductivity sum rule is not changed by a perturbation since
the c-axis kinetic energy remains unchanged in many cases. 
In some notion of interlayer coupling theory,
superconductivity is related to the kinetic energy; namely, it might
be kinetic energy-driven.\cite{chakravarty} 
In this respect the changes in the kinetic
energy and in the c-axis sum rule suggest fundamental information on
superconductivity. Therefore, the $c$-axis
magnetic field is more interesting because it gives 
rise to a shift in the quasiparticle energy spectrum, and serves as 
a direct pair breaker.

When a $c$-axis magnetic field $(H)$ is above $H_{c1}$ and below $H_{c2}$
in a magnitude, where $H_{c1}\; (H_{c2})$ is the lower (upper) critical
field, and the anisotropy of the Dirac cone $\alpha_{D}=v_{F}/v_{G}$
is much larger than one, where $v_{F}\; (v_{G})$ is the Fermi (gap) velocity,
the semiclassical approach\cite{volovik,kubert,vekhter} can be applied. 
Since $H_{c1}\ll H\ll H_{c2}$, vortices are still well separated and
it is 
an extended quasiparticle state that dominately determines the properties
of the in-plane dynamics for a $d$-wave superconductor. 
For $\alpha_{D}\simeq1$,
quantum mechanical effect becomes important;\cite{franz}
however, this is not the case
for the high-$T_{c}$ cuprates, for which $\alpha_{D}\gg1$.

In this paper we consider the effect of a magnetic field
oriented perpendicular to the CuO$_{2}$ planes on the $c$-axis sum rule
and on the corresponding superfluid density. We 
apply a semiclassical approximation to a vortex state in a
two dimensional $d$-wave superconductor for $H_{c1}\ll H\ll H_{c2}$ and
$\alpha_{D}\gg1$, and
include in the calculation only the
energy shift due to the circulating supercurrents around
the vortex cores. We assume that the core occupies only a small part of
the single vortex unit cell, and 
that it is the energy shift of the quasiparticles
outside the cores that is most important. 
We study both the
the case of correlated and uncorrelated pancake vortices, and
find that the $c$-axis sum rule depends on the interlayer correlation of
pancake vortices while details of the intralayer vortex distribution is less
significant.

\section{formalism}

We begin with the Hamiltonian ${\cal H}={\cal H}_{0}+\Hc$, where
${\cal H}_{0}$ describes a $d$-wave superconductivity, and 
$\Hc=\sum_{i\sigma}t_{\perp}\left[c^{+}_{i1\sigma}c_{i2\sigma}+
c^{+}_{i2\sigma}c_{i1\sigma}\right]$
is interlayer coupling;\cite{universal}
therefore, an electron transfers from a site $i$ in plane 1 to
the same site $i$ in plane 2. 
In the semiclassical approximation, a quasiparticle sitting on the 
supercurrent circulating around a votex has an energy shift, the so-called
Doppler shift, and the magnitude of the
supercurrent is inversely proportional to
the distance from the center of the vortex. As illustrated in Fig1.~(a),
for the correlated vortex state a quasiparticle sitting on, say, 
the slow supercurrent
transfers to the next plane without change in momentum
and, therefore, still sits on the slow supercurrent loop as
before. Consequently, we expect that the kinetic energy 
will not change during the interlayer hopping.
Note that the $c$-axis physics is determined by the in-plane
dynamics as we pointed out earlier [See also Eq.~(\ref{kinetic_E})]. 
On the other hand, for the uncorrelated vortex state,
Fig1.~(b), a quasiparticle on the slow supercurrent loop sits on the fast
supercurrent after hopping; therefore, the kinetic energy changes.
The detailed calculation requires the in-plane distribution of
pancake vortices and it will be explained later.

The optical conductivity follows from the appropriate Kubo formula:
\be
\sigma_{c}(\omega)=(i/\omega)\left[
\Pi_{ret}(\omega)-e^{2}d\langle \Hc\rangle\right]\,
\ee
where
$\Pi_{ret}(\omega)$ is the analytic
continuation for
the current-current correlation in Matsubara representation, and 
$\langle\Hc\rangle$ represents the thermal average
of the $c$-axis kinetic energy. 
Here $e$ and $d$ are the electron charge and the interlayer spacing,
respectively.
The superfluid density is detrermined by
Kramers-Kronig relation: $\rho_{s}=4\pi\lim_{\omega\rightarrow0}
\left[\omega{\mbox {Im}}\sigma_{c}(\omega)\right]$.
In terms of the real part of the conductivity $\sigma_{1c}(\omega)$, 
the $c$-axis conductivity sum rule reads:
\be
\rho_{s}=\Delta N
-4\pi e^{2}d
\left[\langle \Hc\rangle^{s}-\langle \Hc\rangle^{n}\right]\;,
\ee
where 
$\Delta N=8\int^{\omega_{c}}_{0^{+}}{}d\omega\left[\sigma^{n}_{1c}(\omega)
-\sigma^{s}_{1c}(\omega)\right]$, and
the superscripts $n$ and $s$ denote the normal and superconducting
state, respectively.  
In the integral defining $\Delta N$, the upper limit is sufficiently large
so as to include all optical transitions of importance to the problem but
not interband effects. 
The kinetic energy change $4\pi e^{2}d
\left[\langle \Hc\rangle^{s}-\langle \Hc\rangle^{n}\right]$ 
will be denoted as $\Delta K$. Therefore, in the presence of a magnetic field
$(H)$, the sum rule becomes $\rho_{s}(H)=\Delta N(H)-\Delta K(H)$, and
it can be written as
\be
{\Delta N(H)\over\rho_{s}(H)}\simeq 1+{\Delta K(0)\over\rho_{s}(0)}
\left[1-{\delta\rho_{s}\over\rho_{s}(0)}\right]
+{\delta K_{s}\over\rho_{s}(0)}\;
\label{sumH}
\ee
where $\rho_{s}(H)=\rho_{s}(0)+\delta\rho_{s}$ and 
$\delta K_{s}=\Delta K(H)-\Delta K(0)$. 
As can be easily seen, the sum rule change due to the field depends 
not only on $\delta K_{s}$ but also on $\delta\rho_{s}$.

\section{magnetic field effects on the kinetic energy and the sum rule}

It is easy to understand that the $c$-axis magnetic field
has no effect on the interlayer hopping for a Fermi liquid in 
the overdoped regime
while the absence of a well-established theory of the pseudogap state 
would seemingly
cause ambiguity in the validity of our approachi for the pseudogap
regime. However, recently
it has been observed\cite{krasnov} that the pseudogap is insensitive to
the magnetic field. This implies that even for the underdoped cuprates,
a semiclassical approach may be applicable to the sum rule calculation
and the kinetic energy change is determined only by the field effect
on the superconducting state; namely, the Doppler shift. 
Here, we would like to emphasize that
we do not calculate the zero field kinetic energy difference $\Delta K(0)$
for the pseudogap state because a model for the pseudogap
would be required to do so;
instead, we base our analysis on experimental observations\cite{basov,krasnov}
for this case. 
We also point out that effects of the Doppler shift
are considered
only on the superconducting state not on the pseudogap state.
\cite{com1}

Let us first calculate the kinetic energy change $\delta K_{s}$. 
In the semicalssical approximation,
the fermionic Matsubara frequencies $i\omega_{n}$ in the 
superconducting $(2\times2)$ Green function
${\hat G}({\bf k},i\omega_{n})$ are shifted by 
${\bf v}_{s}({\bf r})\cdot{\bf k}$, where ${\bf v}_{s}({\bf r})$ is the
superfluid velocity at ${\bf r}$ in a plane and ${\bf k}$ the quasiparticle
momentum,
that is, the $(2\times2)$ Green function in Nambu space
is ${\hat G}({\bf k},i\omega_{n}-{\bf v}_{s}({\bf r})\cdot{\bf k})$, and
an average over the vortex unit cell is to be carried out.
Denoting ${\bar\epsilon}={\bf v}_{s}({\bf r})\cdot{\bf k}$, 
the $(2\times2)$ Green function is given by
\be
{\hat G}({\bf k},i\omega_{n}-{\bar\epsilon})=
{{(i\omega_{n}-{\bar\epsilon}){\hat\tau}_{0}
+\Delta_{\bf k}{\hat\tau}_{1}+\xi_{\bf k}{\hat\tau}_{3}}
\over
{(i\omega_{n}-{\bar\epsilon})^{2}-\xi^{2}_{\bf k}
-\Delta^{2}_{\bf k}}}\;.
\ee
Here $\Delta_{\bf k}$ is a $d$-wave superconducting gap such that
the quasiparticle energy 
$E_{\bf k}=\sqrt{\xi^{2}_{\bf k}+\Delta^{2}_{\bf k}}$ in zero field, and
${\hat \tau}_{i}$'s the Pauli matrices in spin space.
In terms of ${\hat G}({\bf k},i\omega_{n}-{\bar\epsilon})$, 
the $c$-axis kinetic energy in the vortex
state is given at position ${\bf r}_{i}$ by
$K_{s}({\bf r}_{1},{\bf r}_{2})=\sum_{\bf k}
K_{s}({\bf k};{\bar\epsilon}_{1},{\bar\epsilon}_{2})$, where
${\bar\epsilon}_{i}={\bf v}_{si}({\bf r}_{i})\cdot{\bf k}$, with
\be
K_{s}({\bf k};{\bar\epsilon}_{1},{\bar\epsilon}_{2})=
{{\cal C}T\over4}\sum_{\omega_{n}}\mbox{Tr}\left[
{\hat\tau}_{3}{\hat G}({\bf k},i\omega-{\bar\epsilon}_{1})
{\hat\tau}_{3}{\hat G}({\bf k},i\omega-{\bar\epsilon}_{2})\right]\;,
\label{kinetic_E}
\ee
where ${\cal C}=32\pi e^{2}dt^{2}_{\perp}$. The total kinetic energy is
obtained after $K_{s}({\bf r})$ is appropriately averaged over a vortex unit
cell of radius $R$ \ie
$K_{s}(H)=(1/A)\int_{r\leq R} d{\bf r}K^{s}({\bf r})$ 
with $A=\pi R^{2}$ the area
of the unit cell. Actually, the correct averaging procedure depends on whether
or not the pancake vortices from one plane to the next are correlated or
are completely uncorrelated. For the correlated case 
${\bf v}_{s1}({\bf r}_{1})=
{\bf v}_{s2}({\bf r}_{2})$ and there is a simple space average over 
${\bf r}$ while in the uncorrelated case we need to introduce two separate
uncorrelated space variable ${\bf v}_{s1}({\bf r}_{1})$ and
${\bf v}_{s2}({\bf r}_{2})$ and take independent averages over ${\bf r}_{1}$
and ${\bf r}_{2}$ \ie
$(1/A^{2})\int_{r_{1}\leq R} d{\bf r}_{1}\int_{r_{2}\leq R} d{\bf r}_{2}$.

We consider the 
case of uncorrelated pancakes distributed randomly in each plane
so that ${\bf v}_{s1}({\bf r}_{1})\cdot{\bf k}$
and ${\bf v}_{s2}({\bf r}_{2})\cdot{\bf k}$ are to be averaged independently
and separately. In a $d$-wave superconductor it is the nodal quasiparticles
that dominate the low $T$ or the small $H$ response. We take 
$T\rightarrow0$ and work with the difference in kinetic energy with
field on and off \ie $K^{s}(H)-K^{s}(0)$. 
Note that this quantity is determined only by the field effect on 
the superconducting 
state in our consideration.
We then apply a nodal approximation.
In carrying out the
sum over ${\bf k}$ we make a coordinate transformation to ${\bf p}$ with the
nodal point ${\bf k}_{n}$ on the Fermi surface playing
the role of the origin for ${\bf p}$. We then integrate over a nodal
region and the integration
over ${\bf p}$ should not be sensitive to the region of integration provided
$|{\bf p}|$ is taken to be of order $p_{0}\sim{\cal O}(\Delta_{0})$.
With these simplification we can introduce a $\delta$-function 
$\delta\left(\epsilon_{1}-{\bf v}_{s1}({\bf r}_{1})\cdot{\bf k}_{n}\right)$
into the expression for $K_{s}(H)-K_{s}(0)$ and define
\be
{\cal P}(\epsilon_{1})={1\over A}\int d{\bf r}_{1}
\delta\left(\epsilon_{1}-{\bf v}_{s1}({\bf r}_{1})\cdot{\bf k}_{n}\right)\;,
\ee
to obtain
\bea
K_{s}(H)-K_{s}(0)=&&\int^{\infty}_{-\infty}d\epsilon_{1}d\epsilon_{2}
{\cal P}(\epsilon_{1}){\cal P}(\epsilon_{2})
\nonumber\\
&&\times\left[\Psi(\epsilon_{1},\epsilon_{2})-
\Psi(0,0)\right]\;,
\label{ke_un}
\eea
where $\Psi(\epsilon_{1},\epsilon_{2})=\sum_{\bf k}
K_{s}({\bf k};\epsilon_{1},\epsilon_{2})$. 
The matrix sum in $\Psi(\epsilon_{1},\epsilon_{2})$ can be done to get
\bea
\Psi(\epsilon_{1},\epsilon_{2})=&&{{\cal C}\over2}T
\sum_{\bf k}\sum_{\omega_{n}}\Biggl[ G({\bf k},i\omega_{n}-\epsilon_{1})
G({\bf k},i\omega_{n}-\epsilon_{2})
\nonumber\\
&&-F({\bf k},i\omega_{n}-\epsilon_{1})
F({\bf k},i\omega_{n}-\epsilon_{2})\Biggr]\;,
\eea
with $G({\bf k},i\omega_{n})$ and $F({\bf k},i\omega_{n})$
the superconducting state ordinary and anomalous amplitude,
respectively.
For simplicity we will only treat the pure case, and 
after some algebra we derive $\Psi(\epsilon_{1},\epsilon_{2})$ as follows:
\bea
\Psi(\epsilon_{1},\epsilon_{2})=&&{{\cal C}\over4}\sum_{\bf k}
\Biggl\{
{\xi^{2}_{\bf k}\over E^{2}_{\bf k}}\left[
{f_{1+}-f_{2+}\over\epsilon_{12}}+{f_{1-}-f_{2-}\over\epsilon_{12}}
\right]
\nonumber\\
&&+
{\Delta^{2}_{\bf k}\over E^{2}_{\bf k}}\left[
{f_{1+}-f_{2-}\over\epsilon_{12}+2E_{\bf k}}
+{f_{1-}-f_{2+}\over\epsilon_{12}-2E_{\bf k}}
\right]\Biggr\}\;,
\label{psi12}
\eea
where $f_{i\pm}=f(\epsilon_{i}\pm E_{\bf k})$ is the Fermi function, and
$\epsilon_{12}=\epsilon_{1}-\epsilon_{2}$.
Note the symmetry properties:
$\Psi(\epsilon_{1},\epsilon_{2})=\Psi(\epsilon_{2},\epsilon_{1})=
\Psi(-\epsilon_{1},-\epsilon_{2})$ and $\Psi(-\epsilon_{1},\epsilon_{2})=
\Psi(\epsilon_{1},-\epsilon_{2})$. 

The limit $\epsilon_{1}\rightarrow\epsilon_{2}$
is of interest
because it enters the correlated pancake case for which 
${\bf v}_{s1}({\bf r}_{1})\cdot{\bf k}={\bf v}_{s2}({\bf r}_{2})\cdot{\bf k}$.
In this case
the average for the kinetic energy over a single ${\cal P}(\epsilon)$
is 
\be
K_{s}(H)-K_{s}(0)=\int^{\infty}_{-\infty}d\epsilon{\cal P}(\epsilon)
\left[\Psi(\epsilon,\epsilon)-\Psi(0,0)\right]\;,
\label{ke_cor}
\ee
where
\bea
\Psi(\epsilon,\epsilon)=&&{{\cal C}\over4}
\sum_{\bf k}\Biggl\{
{\xi^{2}_{\bf k}\over E^{2}_{\bf k}}
\left[{\partial f(z)\over\partial z}\Big|_{\epsilon+E_{\bf k}}
+{\partial f(z)\over\partial z}\Big|_{\epsilon-E_{\bf k}}\right]
\nonumber\\
&&+{\Delta^{2}_{\bf k}\over E^{2}_{\bf k}}
\left[{f(\epsilon+E_{\bf k})-f(\epsilon-E_{\bf k})\over E_{\bf k}}
\right]\Biggr\}\;.
\eea
In the nodal coordinate system $E_{\bf k}=\sqrt{p^{2}_{1}+p^{2}_{2}}=p$ and
$\sum_{\bf k}=\sum_{\mbox{node}}{\cal J}\int pdpd{\vartheta}$,
where
${\vartheta}$ is a polar angle and 
${\cal J}=\left[(2\pi)^{2}v_{F}v_{G}\right]^{-1}$.
At zero temperature
the Fermi function become $\Theta$-function \ie
$f(z)\rightarrow\Theta(-z)$ such that $\Theta(-z)=0$ for $z>0$ and
otherwise $\Theta(-z)=1$. Substututing $\Psi(\epsilon,\epsilon)$ into
Eq.~(\ref{ke_cor}), we obtain
\bea
K^{s}(H)&&-K^{s}(0)
=2{\cal C}
\int^{\infty}_{0}d\epsilon{\cal P}(\epsilon){\cal J}
\int pdpd{\vartheta}
\nonumber\\
\times&&\Biggl\{{p^{2}_{1}\over p^{2}}\Bigl[-\delta(\epsilon+p)-
\delta(\epsilon-p)\Bigr]
\nonumber\\
+&&\Bigl[\Theta(-\epsilon-p)-\Theta(-\epsilon+p)+1\Bigr]
{p^{2}_{2}\over p^{3}}\Biggr\}\;.
\label{ke_cor1}
\eea
The first $\delta$-function and $\Theta$-function in Eq.~(\ref{ke_cor1})
do not
survive as they give zero contribution.
The first integral in Eq.~(\ref{ke_cor1})
over $p$ gives a contribution proportional to $\epsilon$. The second
integral is limited to the range $p\in(\epsilon,p_{0})$, but its range
can be extended to $(0,p_{0})$ by including the contribution of the first
integral. This combined contribution cancels the last integral
in Eq.~(\ref{ke_cor1}) to give zero.
As pictorially illustrated in Fig1.~(a), we analytically showed that
$K_{s}(H)=K_{s}(0)$ \ie
{\it the out-of-plane magnetic field has no effect on the $c$-axis kinetic
energy regardless of the form of} ${\cal P}(\epsilon)$
{\it in the correlated vortex state}. 
This does not mean, however, that there is no change in the sum
rule given by Eq.~(\ref{sumH}). Instead it means that 
the sum rule is modified only by a possible change
in superfluid density. 
The calculation of $\delta\rho_{s}$ is similar to the above
[See later explanation].
It turns out that
$
\rho_{s}(H)-\rho_{s}(0)=4\pi{\cal C}{\cal J}
\int^{\infty}_{0}d\epsilon\;\epsilon{\cal P}(\epsilon)
$, where $\rho_{s}(0)={\cal C}N(0)/2$ with 
the density of states at the Fermi level $N(0)$, and
we obtain
$\delta\rho_{s}/\rho_{s}(0)=-(2/\Delta_{0})
\int^{\infty}_{0}d\epsilon\;\epsilon{\cal P}(\epsilon)$.

Recently, Vekhter \etal\cite{vekhter4} have given analytic expressions
for different models of the 
vortex distribution function ${\cal P}(\epsilon)$.
They consider
a Gaussian ${\cal P}_{G}(\epsilon)$ and two possible vortex liquid model
${\cal P}_{L1}(\epsilon)$ and ${\cal P}_{L2}(\epsilon)$:
\bea
{\cal P}_{G}(\epsilon)=&&
{1\over\sqrt{\pi}E_{H}}\exp\left(-{\epsilon^{2}\over E^{2}_{H}}
\right)
\\
{\cal P}_{L1}(\epsilon)=&&
{1\over2}{E^{2}_{H}\over(E^{2}_{H}+\epsilon^{2})^{(3/2)}}
\\
{\cal P}_{L2}(\epsilon)=&&
{1\over{\pi E_{H}}}\Biggl[
\arccos\left({1\over\sqrt{(2\epsilon/E_{H})^{2}+1}}\right)
\nonumber\\
&&\times
\left({E^{3}_{H}\over\epsilon^{3}}+{3E^{5}_{H}\over4\epsilon^{5}}\right)
-{3E^{4}_{H}\over2\epsilon^{4}}\Biggr]\;.
\eea
The magnetic energy $E_{H}$ is given by $v_{F}/(2R)$, where 
the radius of the vortex unit cell $R=\sqrt{\Phi_{0}/\pi H}$
with $\Phi_{0}$ the flux quantum. Using
the model ${\cal P}(\epsilon)$, we obtain 
$\delta\rho_{s}/\rho_{s}(0)=-\alpha_{1}E_{H}/\Delta_{0}$.
where $\alpha_{1}$ is $1/\sqrt{\pi}$ for $G$,
$1$ for $L1$, and $0.84$ for $L2$.
Note that all distributions
give the same linear dependence on $E_{H}/\Delta_{0}$
with slightly different coefficients, and $H$ reduces
the superfluid density as we expect.

Our main results so far 
are that in the correlated vortex case there is no change
in kinetic energy on application of a magnetic field $H$ 
perpendicular to the layers, independent of
the vortex distribution function
${\cal P}(\epsilon)$.
In all cases considered,
the superfluid density is reduced by the magnetic field,
and $\delta\rho_{s}/\rho_{s}(0)$ is linearly proportional to
$E_{H}/\Delta_{0}$ with a coefficient of order $1$.
The form of ${\cal P}(\epsilon)$ determines
the exact value of the coefficient. We consider
two interesting regimes of doping for the correlated vortex case 
$(\delta K_{s}=0)$. 
For the overdoped case we assumed that
the in-plane motion can be described by a Fermi liquid. In this
case, it has been found previously\cite{kim1} that
in zero field
$\Delta N(0)/\rho_{s}(0)=1$ \ie the $c$-axis sum rule
is conventional, which is consistent with experimental
observations.\cite{basov} It arises
because $\Delta K(0)=0$ in this instance.
Therefore, it follows
from Eq.~(\ref{sumH})
that $\Delta N(H)/\rho_{s}(H)=1$ also \ie
there is no change in the sum rule induced by the external field.

As we mentioned earlier,
we do not rely on a specific theory to calculate the sum rule for
the underdope regime.
Instead we use experimental observations\cite{basov} on the underdoped 
cuprates.
In this case the zero field sum rule is observed to be about a half; namely,
$\Delta N(0)/\rho_{s}(0)\simeq 1/2$, which also follows directly in the
preformed pair model for the pseudogap. 
This means that
$\Delta K(0)/\rho_{s}(0)=-1/2$ in Eq.~(\ref{sumH}).
When we calculate $\delta K_{s}$,
we take into account only effects of the Doppler shift due to 
the magnetic field,
again, based on experimental results.\cite{com2}
Consequently, we obtain
$
{\Delta N(H)/\rho_{s}(H)}=1/2
+{\delta\rho_{s}/2\rho_{s}(0)}=
{1/2}-\alpha_{1}E_{H}/2\Delta_{0}
$, where $\alpha_{1}=1/\sqrt{\pi}$ for $G$, $1$ for $L1$, and
$0.84$ for $L2$.
The sum rule originally equal
to $1/2$ is
further reduced by the magnetic field.

Now, we return to Eq.~(\ref{ke_un}) for the uncorrelated vortex state 
with Eq.~(\ref{psi12})
defining $\Psi(\epsilon_{1},\epsilon_{2})$. 
In this case $K^{s}(H)-K^{s}(0)$ becomes
\bea
&&K^{s}(H)-K^{s}(0)
\nonumber\\
=&&
2\int^{\infty}_{0}d\epsilon_{1}d\epsilon_{2}{\cal P}(\epsilon_{1})
{\cal P}(\epsilon_{2})\left[\Psi(\epsilon_{1},\epsilon_{2})-
\Psi(0,0)\right]
\nonumber\\
+&&2\int^{\infty}_{0}d\epsilon_{1}d\epsilon_{2}{\cal P}(\epsilon_{1})
{\cal P}(\epsilon_{2})\left[\Psi(\epsilon_{1},-\epsilon_{2})-
\Psi(0,0)\right]\;.
\eea
On application of the nodal approximation to the difference
$\Psi(\epsilon_{1},\epsilon_{2})-\Psi(0,0)$ in the limit $T\rightarrow0$,
we obtain
\bea
\delta K_{s}=&&{{\cal C}\over2}{N(0)\over\Delta_{0}}
\int^{\infty}_{0}d\epsilon_{1}d\epsilon_{2}{\cal P}(\epsilon_{1})
{\cal P}(\epsilon_{2})
\nonumber\\
&&\times\left\{{\epsilon_{1}\epsilon_{2}\over(\epsilon_{1}+\epsilon_{2})}
+{1\over2}(\epsilon_{1}+\epsilon_{2})
\ln\left|{\epsilon_{1}-\epsilon_{2}\over
\epsilon_{1}+\epsilon_{2}}\right|\right\}\;,
\eea
which is of order $E_{H}/\Delta_{0}$, and numerical calculations give 
$\delta K^{s}/\rho_{s}(0)=-\alpha_{2}E_{H}/\Delta_{0}$,
where $\alpha_{2}$ is $0.123$ for $G$, $0.135$ for $L1$,
and $0.102$ for $L2$.
We also derive the change in superfluid density which follows in
a similar fashion; namely,
$\delta\rho_{s}=\int^{\infty}_{-\infty}d\epsilon_{1}d\epsilon_{2}
{\cal P}(\epsilon_{1}){\cal P}(\epsilon_{2})
\left[\chi(\epsilon_{1},\epsilon_{2})-
\chi(0,0)\right]$, 
where 
\bea
\chi(\epsilon_{1},\epsilon_{2})=&&{{\cal C}\over4}\sum_{\bf k}
{\Delta^{2}_{\bf k}\over E^{2}_{\bf k}}\Biggl[
{f_{1+}-f_{2+}\over\epsilon_{12}}+{f_{1-}-f_{2-}\over\epsilon_{12}}
\nonumber\\
&&-{f_{1+}-f_{2-}\over\epsilon_{12}+2E_{\bf k}}
-{f_{1-}-f_{2+}\over\epsilon_{12}-2E_{\bf k}}
\Biggr]\;.
\eea
It can be shown that the superfluid density reduction becomes
\bea
\delta\rho_{s}=&&-{{\cal C}\over2}{N(0)\over\Delta_{0}}
\int^{\infty}_{0}d\epsilon_{1}d\epsilon_{2}{\cal P}(\epsilon_{1})
{\cal P}(\epsilon_{2})
\nonumber\\
&&\times
\left\{{3(\epsilon_{1}+\epsilon_{2})^{2}+\epsilon^{2}_{1}+\epsilon^{2}_{2}
\over2(\epsilon_{1}+\epsilon_{2})}+{(\epsilon_{1}+\epsilon_{2})\over2}
\ln\left|{\epsilon_{1}-\epsilon_{2}\over
\epsilon_{1}+\epsilon_{2}}\right|\right\}\;,
\eea
and 
$\delta\rho_{s}/\rho_{s}(0)=-\alpha_{3}E_{H}/\Delta_{0}$,
where $\alpha_{3}$ is $0.34$ for $G$, $0.72$ for $L1$, and
$0.63$ for $L2$. For correlated vortex, we used $\chi(\epsilon,\epsilon)$
with $\chi(0,0)=\rho_{s}(0)$.

The sum rule of the uncorrelated vortex state reads
$
{\Delta N(H)/\rho_{s}(H)}\simeq1-\alpha_{2}(E_{H}/\Delta_{0})$
for the overdoped, and 
${\Delta N(H)/\rho_{s}(H)}\simeq
1/2-(\alpha_{2}+\alpha_{3}/2)(E_{H}/\Delta_{0})$ for the underdoped regime.
In both cases the sum rule is reduced by the presence of a magnetic field. 
In the overdoped case the reference is, however,
$1$ while for the underdoped case it is $1/2$, and the coefficient of 
the reduction,
which goes like
$E_{H}/\Delta_{0}$, is larger for the underdoped case as compared with
the overdoped case.
Note that the magnetic energy $E_{H}$ goes like
$\sqrt{H}$ in all cases.

\section{conclusions}

For {\it correlated} pancake vortices with
an in-plane Fermi liquid and coherent $c$-axis coupling,
there is no change in the $c$-axis optical sum rule when 
a magnetic field is applied perpendicular to the layers of 
a $d$-wave superconductor. The sum rule keeps its conventional 
value of one. At the same time the superfluid density in the 
$c$-direction decreases by an amount proportional to the square 
root of the magnitude of the external magnetic field $(\sqrt{H})$. 
If, however, the in-plane dynamics is unconventional, 
for example, it can be described by a pseudogap above $T_{c}$,
the sum rule, which is equal to $1/2$ when $H=0$, gets reduced 
linearly in $\sqrt{H}$ as does the $c$-axis 
superfluid density.
For {\it uncorrelated} vortices, there is always a 
reduction in the sum rule which is also proportional to $\sqrt{H}$.
For the in-plane
Fermi liquid case the sum rule is reduced below one, 
while for the  unconventional case with a pseudogap it is reduced below $1/2$. 
In both cases a reduction in the $c$-axis superfluid density accompanies
the sum rule reduction.

\begin{center}
{\bf AKNOWLEDGMENTS}
\end{center}

Research was supported in part by the National Sciences and Engineering
Research Council of Canada (NSERC) and by the Canadian Institute for
Advanced Research (CIAR).

\end{multicols}
\begin{figure}
\caption{ Pictorial illustration for a correlated (a) and an uncorrelated (b)
vortex state. $v_{s}$ is the supervelocity and the gray tubes are vortex
cores.
}
\end{figure}


\begin{references}
\bibitem{com2} This is in accord with the preformed pair model. One could
conceive of more complicated origin of the pseudogap in which its
effect persists below $T_{c}$ as distinct from superconductivity and so
modifies the supercurrents in a way not specifically treated here.
\end{references}
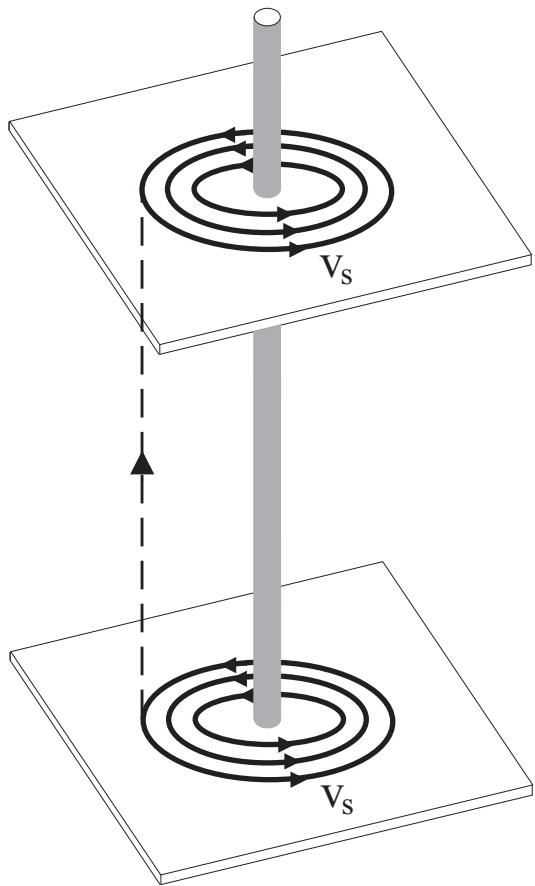
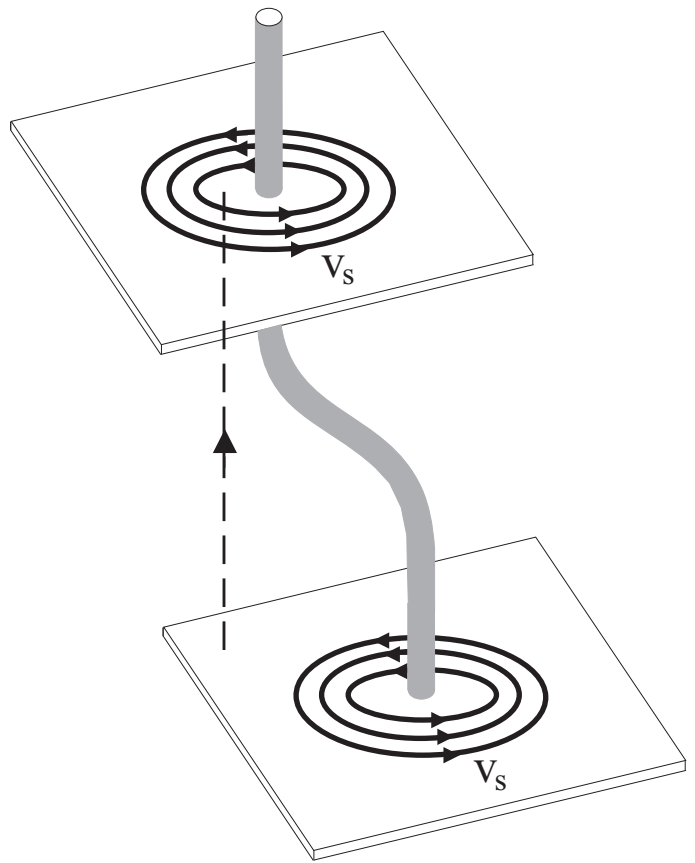
\end{document}